\documentclass[draft,jgrga]{agutex}
\usepackage{amssymb}
\usepackage{graphicx}
\setkeys{Gin}{draft=false}

\titlerunninghead{FIREBALL SPECTRUM}
\authorrunninghead{OBENBERGER ET AL.}
\authoraddr{Corresponding author: K.S. Obenberger,
(kso1987@unm.edu)}

\begin{document}

\title{Dynamic Radio Spectra from two Fireballs}

\authors{K.S. Obenberger,\altaffilmark{1} G.B. Taylor,\altaffilmark{1}  C. S. Lin,\altaffilmark{2}  J. Dowell,\altaffilmark{1}  F.K. Schinzel,\altaffilmark{1} K. Stovall,\altaffilmark{1}}
\altaffiltext{1}{Department of Physics and Astronomy, University of New Mexico, Albuquerque, NM 87131}
\altaffiltext{2}{Institute For Scientific Research, Boston College, Chestnut Hill, MA 02467}

%

\begin{abstract} We present dynamic spectra from the LWA1 telescope of two large meteors (fireballs) observed to emit between 37 and 54 MHz. These spectra show the first ever recorded broadband measurements of this newly discovered VHF emission. The spectra show that the emission is smooth and steep, getting very bright at lower frequencies. We suggest that this signal is possibly emission of Langmuir waves and that these waves could be excited by a weak electron beam within the trail. The spectra of one fireball displays broadband temporal frequency sweeps. We suggest that these sweeps are evidence of individual expanding clumps of emitting plasma. While some of these proposed clumps may have formed at the very beginning of the fireball event, others must have formed seconds after the initial event.

\end{abstract}

\begin{article}

\section{Introduction}


Generally speaking, fireballs are meteors that appear optically brighter than the planet Venus (magnitude $-$4) , and occur from meteoroids larger than $\sim$ 100 grams. Such large meteors are considered rare given that an observable area of $10^{6}$ km$^{2}$ has a rate of occurrence of $\sim$ 200 per year \citep{Halliday96}. Several fish eye camera observatories currently operate year round, finding hundreds of fireballs a month. For instance the NASA All Sky Fireball Network (http://fireballs.ndc.nasa.gov) operates 15 cameras in 8 different states including two in New Mexico and 3 in Arizona. These cameras are situated such that they can triangulate the position of the fireballs, and thereby get the velocities and orbital parameters. 

Similarly the First Station of the Long Wavelength Array (LWA1; \citet{Ellingson13}; \citet{Taylor12}), located in central New Mexico, creates images of the entire sky. However these images probe the low end of the electromagnetic spectrum (10 - 88 MHz) in the high frequency (HF) and very high frequency (VHF) bands. In 2014 the LWA1 was used to serendipitously discover that meteors, specifically fireballs, radiate non-thermal emission in this frequency range \citep{Obenberger142}. These radio transients were identified as fireballs in part because the two Fireball Network cameras in New Mexico are close enough to the LWA1 to see some of the same events. 

Little is known about the radio properties of fireballs. In \citet{Obenberger142} we used polarization, light curves, sky position, meteor shower correlation, and a limited amount of spectral information to show that the 49 radio transients we detected were most likely due to non-thermal radio emission from fireballs. These fireballs were detected at 25.6, 30, and 38 MHz, and it was noted that those at 25.6 MHz were considerably brighter, suggesting that the emission had a steep spectrum. The light curves of these fireballs were all very similar, displaying a fast linear rise followed by a slower exponential decay, with a total duration of 10 to 200 seconds. Such long durations are unusual for atmospheric radio transients at these frequencies; this fact alone sets this phenomenon apart from other sources. Radio frequency (RF) emission from electrical discharges such as lightning are short (10s of ms) \citep{Warwick79}, therefore the two are most likely unrelated. Similarly the fireball emission does not appear to be related to the other known meteor RF phenomenon, namely the sub microsecond EMPs caused by hypervelocity impacts \citep{Close13,Close10} and the short pulses of very low frequency (VLF) and extremely low frequency (ELF) radio emission observed to come from large fireballs and bolides \citep{Guha12,Keay80,Beech95}. Finally the fact that the gyro-frequency of the geomagnetic field is $\sim$1.5 MHz, an order of magnitude less than the observed frequencies, indicates that the emission is not cyclotron.

 A possible candidate for the emission is the radiation of Langmuir waves from the meteor plasma trail. These waves could possibly be driven by an electric and/or turbulent interaction with the ionosphere. The observed emission would be at the local plasma frequency, $f_{p} = \sqrt{ n_{e} e^{2} / \pi m_{e}}$, where $n_{e}$ is the electron density, $e$ is the charge of the electron, and $m_{e}$ is the mass of the electron. This and the rest of the equations in the paper are given in cgs units. Since the meteor trail would contain a gradient in the electron densities, such emission would be a probe into the profile of the trail and give clues to how the trail evolves. 

The initial fireball emission discovery was made using images from the Prototype All-Sky Imager (PASI; \citet{Obenberger15}), a backend correlator of the LWA1 telescope. PASI creates narrowband (75 kHz) all-sky images, integrated for 5 seconds; and the center frequency can be tuned anywhere between 10 and 88 MHz. The images are permanently stored in an archive which now contains over 11 million images ($>$ 15,000 hours). These images have been searched in the manner described in \citet{Obenberger142} and have resulted in the detection of 101 transients, the vast majority (if not all) of which are fireballs. 

We also observe with the LWA1 phased array mode, where the antennas are delayed and summed to form a pencil beam. Since this is done digitally we can point multiple beams at one time. During PASI operation we fix three beams on the sky around zenith, covering a narrow region ($\sim 50$ deg$^{2}$) of the sky. These observations had the primary objective of finding dispersed transient events, and in addition they could be used to capture low frequency spectra from fireballs. The beams provide 36 MHz of bandwidth coverage with high frequency and time resolution, and since we can readily detect fireballs with PASI, if one occurs within a beam we can retrieve the broadband spectra. This campaign has been successful, with the recording of two dynamic spectra from fireballs. 

This paper presents radio dynamic spectra of these two fireballs, and is organized as follows: \S2 describes the observations made. \S3 provides analysis of the image data. \S4 provides analysis of the dynamic spectra. \S5 describes why the observations cannot be due to reflections of celestial sources. \S6 provides a basic model for the emission. Finally \S7 presents our conclusions.

\section{Observations} In October of 2013, we began an observing campaign to search for transients with LWA1 beam formed data. This mode of operation uses three beams centered around zenith at azimuths of 60$^{\circ}$, 180$^{\circ}$,  and 240$^{\circ}$ each at an elevation of 87$^{\circ}$, and runs simultaneously with the PASI all-sky imaging mode described in \citet{Obenberger15}. For simplicity, these beams will hereafter be referred to as B1 (60$^{\circ}$), B2 (180$^{\circ}$), and B3 (240$^{\circ}$).

Each beam has two tunings centered at 45.45 MHz and 65.05 MHz, and each tuning has 19.6 MHz of bandwidth, only 18 MHz of which are usable due to the digital filter roll off. These center frequencies were selected to avoid the video carrier for analog TV channel 2 at 55.25 MHz, which is one of the strongest radio frequency interference (RFI) emitters in the LWA1 frequency range. This channel has actually been used by LWA1 scientists to study meteors using forward scatter \citep{Helmboldt14}.  

Rather than saving the raw voltage time series, we use the spectrometer mode, provided through the data recording subsystem, which channelizes, integrates, and records the data in real time at a much lower data rate. For these observations we selected a mode that provides 1024 19.14 kHz wide channels and 40 ms integration steps. Due to limited computational resources, only the XX and YY auto correlations are computed, which allows us to calculate the Stokes I and Q polarizations. However, sine we do not compute the XY and YX cross correlations we cannot calculate Stokes U and V. The dynamic spectra are recorded at a data rate of 1.37 GB hr$^{-1}$ beam$^{-1}$ and are copied to the LWA1 Data Archive. This project has accumulated over 5,600 beam hours to date.

Fireballs are detected and located using PASI and the image subtraction pipeline described in \citet{Obenberger15}. When a fireball occurs within one of the three beams, the corresponding data can be downloaded from the archive and analyzed. So far this has occurred twice, with both events occurring during the Orionid meteor shower. The first was on October 17, 2014 at 7:21 UT (hereafter referred to as FB1) and the second on October 26, 2014 at 12:58 UT (hereafter referred to as FB2). FB1 was also caught by the NASA All-Sky Fireball Network and was measured to have a velocity of 52 km/s with an upper altitude of 100.9 km and lower altitude of 85.9 km. The estimated peak absolute magnitude was $\sim$ -3 resulting in an estimated initial mass of $\sim$ 10 grams.

Without optical spectra or LIDAR it is impossible to determine the temperature of the plasma, especially in the 10s of seconds after the initial optical emission. For typical meteors, the initial temperature of the plasma trail is between 5,000 and 10,000 K \citep{Ceplecha98}, and quickly decreases coming into equilibrium with the ambient surroundings ($\sim$ 200 K) usually within milliseconds up to seconds depending the size of the meteor \citep{Ceplecha98,Borovicka06}. However in the case of optical persistent trains, a potentially related phenomenon, the walls of the plasma tube have been shown through LIDAR measurements to be about 20 - 50 K above the ambient surroundings 3 minutes after ablation and ionization \citep{Chu00}.

\section{PASI Images Analysis}
The transient detection pipeline is straightforward and described in detail in \citet{Obenberger15}. Once detected however, further analysis needs to be conducted to determine whether the transients are indeed fireballs. In the two cases presented in this paper, it is clear from their light curves and resolved image elongation that they match the description of fireballs presented in \citet{Obenberger142}. As can be seen in Figure 1, the light curves display the typical shape of a fast rise and slower decay, and the images of the fireballs show clear elongation. Moreover FB1 was confirmed by the fireball network to have an optical counterpart leaving no doubt that it was indeed a fireball. FB2 occurred 21 minutes before sunrise and therefore could not have been detected by the fireball network.

Figure 1 also shows the proximity of the three beams to the trails of the fireballs. It is clear that not all of the beams will have equal amounts of emission within them, moreover some lack any significant emission. In particular for FB1, PASI shows that B1 did not have any significant emission present, which was confirmed when analyzing the beam data. The fireball network calculated a trajectory for FB1 heading NW, which in the PASI image is to the upper right. Therefore B3 is located near the end of the trail, while B2 is located more towards the middle, which is the brightest region. For FB2, the beams were more equally distributed, with B2 containing the most received power. However since FB2 was not detected by the fireball network, we do not know the trajectory.

Still, we may be able to determine the trajectory of FB2. Both FB1 and FB2, along with many other fireballs, create observable streaks in the radio images. These streaks are similar to the optical counterparts, in that the radio emission starts at one end of the trail and moves to the other end. However it typically takes 10 to 40 seconds for the radio emission to reach the end of the trail, whereas the optical portion typically takes less than 2 seconds. Therefore the radio streak is not displaying the physical formation of the trail. Despite this, in all 10 cases (including FB1) where a fireball was detected by the fireball network and displayed observable elongation in PASI, the emission moves from the upper portion of the trail to the lower portion.  

The radio emission of FB2 streaks to the south east. Therefore, if we assume that the emission starts at the upper region for all fireballs, then we can say that FB2 was headed SE. This method may allow for the LWA1 to determine the trajectory of all observed fireballs. 

Given the fact that the timescale of the upper-to-lower transition is much too long to be tracing the initial formation of the plasma we consider the idea that it may be related to the diffusion of the dense plasma trail into the less dense ionosphere. The diffusion constant is exponentially dependent on the altitude, where the diffusion rate increases at higher altitudes. Therefore if the majority of the trail has a high enough electron density that the plasma frequency is above the observing frequency of PASI, then as the trail expands the plasma frequency of the upper regions will reach the observing frequency first, followed by the lower regions. It should be noted that the diffusion would be further complicated at higher altitudes by geomagnetic field and nonlinear diffusion \citep{Jones91,Dyrud01}. If the emission is indeed coming from radiation of Langmuir waves, then further study may prove useful for studying high elevation diffusion phenomena.

\section{Dynamic Spectra Analysis}

Each dynamic spectra had a model bandpass divided out. This model is based on the responses from the antennas and filters. Once the bandpass is divided out, the background sky needs to be subtracted out. This is done by simply subtracting a smoothing spline fit; spectral features due to RFI were removed in the data that we fit. to the median spectrum from the minute prior to each fireball. This method works well because the spectral and amplitude features of the sky change very little over the period of a few minutes, and this is confirmed by the fact that the sky corrected spectrum is still flat and centered at zero after the fireball emission has ended. 

 In all of the beam data there were several narrow band transmitters that leaked into the beams via the side lobes. Furthermore, both plasma trails reflected numerous additional transmitters. The majority of these frequencies have been masked out in the dynamic spectra. In a few instances the reflections are so bright that they saturate the beam formed output, and all other frequencies at those times drop significantly in power. These times have been masked out as well. It is interesting to note that the duration of the narrow band reflections as a function of frequency is similar to the broad band emission. This would make sense if the emission was coming from Langmuir waves given that the reflections of a transmitter would end once the plasma frequency dropped below the transmitter's frequency.

No significant flux for either FB1 or FB2 was detected in B1. In B3 we did detect FB1, although it contained less flux than B2 as was expected from the PASI images in Figure 1. For FB2, B3 was completely saturated by a very strong reflection lasting the duration of the event.

Figure 2 shows the full dynamic spectra from B2 of FB1 and FB2. Figures 3 and 4 show closeups from B2 of FB1, in both Stokes I and Stokes Q, while Figure 5 shows a similar closeup of B3.

\subsection{Unpolarized Broadband Power Emission}
Qualitative inspection of the dynamic spectra of each fireball revealed that the received power has a steep, smooth dependence in frequency for most of its duration, where the power increases at lower frequencies. This was not the case in FB1 between 10 and 20 s, which contained narrow, linearly polarized frequency sweeps with enhanced emission from 40 to 41, 42 to 44, and 50 to 54 MHz. Figures 3, 4 and 5 show that these regions are linearly polarized with enhanced dark red between 40 and 44 MHz, and enhanced yellow between 50 and 54 MHz. Figure 6 shows the averaged spectra from the times when these sweeps are present and when they are not. The smooth emission from FB1 was analyzed by examining the averaged spectrum between 26 and 34 s, which only exhibited a small number of spectral sweeps. No spectral sweeps were detected in the case of FB2, thus the entire duration of the emission was averaged.



While we accounted for the bandpass of the instrument, there is still an instrumental contribution to the spectra, and there are two components to this. First the full width at half maximum (FWHM) beam size depends on the frequency like: FWHM$\sim c/fD$ where $f$ is the frequency, $c$ is the speed of light, and $D$ is the diameter of the telescope. So if a source is off the center of the beam, the higher frequencies will have a weaker response. Secondly, the meteor is not a point source, rather it is a line with effectively no width. This is because a fireball in this size range would only create a plasma trail tens of meters wide, which would appear on arc second scales from $\sim$ 100 km away, far smaller than the size of the LWA1 beams. Therefore more of the physical source is seen at the lower frequencies, further adding to the spectral shape. 

We modeled these effects by creating 2D Gaussian point spread functions (PSF) for a 100 m diameter dish at 13 frequencies from 20 to 85 MHz. We then simulated an infinitely long meteor with a width that is much smaller than the angular beam size and flat spectrum passing through each PSF at different offsets from center. Since the beam-width goes like $f^{-1}$, the zero angle offset is a perfect power law with index -1. However, as the offset angle increases, the shape deviates from a power law.

We estimate the angle offsets of FB1 and FB2 from the nearest beam (B1 in both cases) by simply using the images in Figure 1. Doing this gives offsets of $\sim 0.75^{\circ}$ and $\sim 2.75^{\circ}$. Dividing out the spectral contribution for these offsets gives a seemingly linear dependence below $\sim$ 50 MHz. Above 50 MHz the spectra of both fireballs flatten to zero. See Figure 7. 

The temporal smoothness of the dynamic spectrum further bolster the argument that the observed fireballs are not mere reflections. Reflections fluctuate in brightness due to both interference of multiple reflection points as well as the changes in shape and orientation of the trail caused by local winds. The dynamic spectra show smoothness on timescales of 40 ms, which is much finer than the previous finest example showing 1 s resolution \citep{Obenberger142}.

. 

\subsection{Polarized Broadband Frequency Sweeps}

The dynamic spectrum of FB1 contains linearly polarized spectral sweeps. Figures 3 and 4 show closeups of these regions in B2, with both Stokes I and Q polarizations, and Figure 5 shows a similar closeup of B3. Given their large bandwidth and periodicity they are certainly not reflected radar, and are most likely an inherent feature of the emission. The sweeps seen in B2 are very defined and have a frequency separation of $\sim$ 1 MHz and a temporal separation of $\sim$ 250 ms, whereas the sweeps seen in B3 are less defined and have a frequency separation of $\sim$ 1.5 MHz and a temporal separation of $\sim$ 900 ms. 

\subsubsection{Polarization}
With only Stokes I and Q we do not get the full polarization picture. However, at least some portion of the sweeps are linearly polarized, and this polarization is the same with both the sweeps seen in B2 and B3. At 43 MHz the sweeps are $\sim 80\%$ -Q, whereas at 51 MHz they are $\sim 50\%$ $+$Q, with the magnitude of Q polarization dropping off significantly on either side of these two frequencies. We find that Faraday rotation cannot account for the severity of the change from -Q to +Q. To show this we take 51 and 43 MHz to be the frequency separation between a 90$^{\circ}$ rotation in polarization. Assuming that the pulse has $\sim 80\%$ polarization at 43 MHz and $\sim 50\%$ at 51 MHz, the angle of rotation between two frequencies is:

\begin{equation}
\gtrsim\Theta = RM c^{2} \left(\frac{1}{f_{1}^{2}} - \frac{1}{f_{2}^{2}}\right),
\end{equation}

\noindent where $RM$ is the commonly used rotation measure, which is given by:

\begin{equation}
RM  = \frac{e^{3}}{2 \pi m^{2} c^{4}} \int_0^d n_e(l) B_{||}(l) dl,
\end{equation}

\noindent and $B_{||}$ is magnitude of the magnetic field parallel to the propagation of the wave. In this case since the fireball was at zenith, $B_{||}$ is just the vertical component of the local geomagnetic field, and is considered constant between 0 and 100 km. Using the National Geophysical Data Center of the National Oceanic and Atmospheric Association (www.ngdc.noaa.gov/geomag-web/) we estimate $B_{||}$ to be 0.4 G. Therefore, assuming a  90$^{\circ}$ polarization angle rotation between 43 and 51 MHz, we calculate an electron column density of $1.2\times10^{12}$ cm$^{-2}$.

While we do not have the exact electron density of the ionosphere surrounding the fireball trail, the event occurred at 01:21 MT, which implies that the electron density of the ionosphere was most likely very low. Indeed, we can limit the density using public ionogram data from the Digital Ionogram Database (DID). At the time of the fireball, the electron density below 200 km for Boulder, CO; Austin, TX; and Melrose, CA was less than $10^{3}$ cm$^{-3}$. Therefore, for the ionosphere to contribute such a large rotation would require a path length of 10,000 km, rendering this scenario impossible.

Next we consider the electron contribution from the fireball trail itself. We assume the pulse originated inside the trail, traveling a distance $d$ through the plasma. Since the lower end of our band is $\sim$ 37 MHz we set an upper limit for the plasma frequency within the distance $d$ to 37 MHz, which corresponds to an electron density of $1.7\times10^{7}$ cm$^{-3}$. At such a density the required distance would be $\sim 760$ m. As can be seen from figure 1 the angular size of FB1 is $\sim 20^{\circ}$, which gives it a projected length of $\sim$ 30 km at $\sim$ 90km, and the fireball network calculates a similar length of 32 km. Therefore the fireball appears to be viewed nearly edge on, where the projected depth of the trail is approximately equal to the actual width of the trail. Therefore it is unlikely that the trail would have expanded to a width of 760 km 10 seconds after impact. Moreover, while a plasma density of 37 MHz is reasonable for a portion of the trail, it is very unlikely that such a density would expand outward for 100s of meters. 

For these reasons the opposite polarity at 43 and 52 MHz appears to be an inherent property of the frequency sweeps, and are no doubt clues to their origin.

\subsubsection{Frequency-Time Dispersion}
At first glance it is difficult to tell if the sweeps in B2 are continuous across the entire band. There appears to be a decrease in brightness between 45 and 50 MHz, moreover there was a significant amount of RFI removed from this portion of the band. Nevertheless, we can see that the sweeps do move through the whole band, an observation that is perhaps more obvious for the sweeps in B3 (Figure 5) than those in B2 (Figure 3). This is also confirmed when we extracted the brightest points from the sweeps below and above the region of decreased emission, and they were noticed to be connected via a power law. We therefore modeled each set of points using a least squares fit of the function:

\begin{equation}
t(f) = \frac{a}{f^{b}} + c,
\end{equation}

\noindent where $t$ is the arrival time of the pulse, $f$ is the frequency and $a$, $b$, and $c$ are all free parameters. The results of the fits are given in Table 1. The values of $b$ ranged from 0.8 to 1.8, with an average value of 1.28, and standard deviation of 0.35. It is interesting to note that all modeled values of $b$ are less than 2. If the dispersion was caused by typical electromagnetic dispersion in a plasma, where the group velocity of a wave is a function of frequency, the measured value for $b$ should always be greater than or equal to 2. This is because within an unmagnetized plasma the group velocity for an electromagnetic wave is:

\begin{equation}
v_{g} = c\sqrt{1 - \frac{f_{p}^{2}}{f^{2}} },
\end{equation}

\noindent where $f_{p} $ is the plasma frequency and $c$ is the speed of light. We are only considering O-mode waves since the gyro-frequency of the geomagnetic field is nearly a factor of 30 less than the observed frequencies and thereby the assumed plasma frequencies. Therefore the upper hybrid frequency is only $\sim20$ kHz greater than the plasma frequency. At this time we cannot rule out the possibility that the plasma emission could contain hybrid waves; further study of the radio emission may shed light on the responsible wave modes. Either way the fact that the hybrid frequency is so close to the plasma frequency indicates that the unmagnetized approximation in Equation 4 is valid for the propagation of X-mode waves as well. Therefore the time of arrival of a pulse generated at a distance $d$ within a plasma can then be calculated as:

\begin{equation}
t(f) = \int_0^d v_g dl = \frac{1}{c}\int_0^d\frac{dl}{\sqrt{1 - \frac{f_{p}^{2}}{f^{2}} }} = \frac{1}{c}\int_0^d\left(1 +  \frac{f_{p}^{2}}{2f^{2}} +  \frac{3f_{p}^{4}}{8f^{4}} +  \frac{5f_{p}^{6}}{16f^{6}} + O(f^{-\delta})\right)dl.
\end{equation}

Comparing Equation 3 and Equation 5 it can be seen that the value for $b$ due to this type of dispersion would be a minimum of 2. Moreover this only occurs when $f \gg f_{p}$, which is certainly not the case for the observed fireballs. Therefore the frequency-time dispersion must have some other explanation.

\section{Reflections from Celestial Continuum Sources?}
Up until this point we have assumed the conclusions of \citet{Obenberger142}, namely that the observed fireballs are emitting. In \citet{Obenberger142} we showed that it was very unlikely that the detected signals associated with fireballs were due to reflections of man-made RFI. However we did not address the possibility of reflections from bright celestial sources (e.g. Taurus A, Virgo A, Cassiopeia A, Cygnus A, and the Galactic Center). Such reflections would not be subject to many of the previously presented arguments against reflections of man-made RFI. In particular these sources are unpolarized and broadband, so they are similar to the fireballs in this respect. Nevertheless, we find that such reflections are highly unlikely for the following reasons.

For one, reflections of celestial sources should also undergo the same sporadic variations seen in reflections of man-made RFI. As described in \S 4.1 the radio counterpart to fireballs are very smooth in time.

Moreover there are only 5 sources brighter than the fireballs we see ($\gtrsim$ 1 kJy at 38 MHz). Therefore the probability that the proper geometry would be present for a specular reflection is very low, certainly much lower than the occurrence of fireballs we observe \citep{Obenberger142}. 

Furthermore the received power from an overdense reflection drops like the inverse square of the distance to the trail \citep{McKinley61,Ceplecha98,Wislez95}. Therefore such reflections should be much dimmer than the sources they are reflecting. However in several cases the received power is 10 to 20\% of the brightest possible reflected source.

\section{The Nature of the Emission}
In this section we provide a simple hyopothesis that is a first approach to explain the emission through radiation of Langmuir waves. Since the cyclotron frequency is over a factor of 30 less than the observed plasma frequencies we ignore the role of the geomagnetic field. It may be that the magnetic field is an important factor for the emissive process, so the following hypothesis should only be considered valid parallel to the magnetic field.

\subsection{Langmuir Wave Generation}

Meteor trails possess the necessary electron densities such that if Langmuir waves were produced they would be within the LWA frequency range \citep{Ceplecha98,McKinley61}. Yet this fact neither explains the occurrence of Langmuir waves nor the mechanism by which they are radiated.

The Langmuir waves could not be left over from the initial ablation and ionization because such waves would be damped out due to the high electron-neutral collision rate. From \citep{Kelley09}, the electron-neutral collision frequency for the lower ionosphere is given by:

\begin{equation}
\gamma_{en} = 5.4 \times 10 ^{-10} n_{n} \sqrt{T_{e}},
\end{equation}

\noindent where $n_{n}$ is the neutral density and $T_{e}$ is the electron temperature. The neutral density at 90 km is 3.9$\times 10^{13}$ cm$^{-3}$ \citep{Kelley09} and the electron temperature has come to equilibrium with the ambient ionosphere at $\sim$ 200 K. Therefore the electron/neutral collision frequency is $\gamma_{en} = 3\times 10^{5}$ s$^{-1}$. The observed plasma frequencies are $\sim$ 100 times greater than the collision frequency, allowing plasma waves to be driven and amplified, provided the driving mechanism has a growth rate greater than the collision frequency.

In comparison Type II and III solar radio bursts are thought to be the product of Langmuir wave radiation \citep{Reid14,Melrose87}. Electron beams are well known to be efficient drivers of Langmuir waves through the bump-in-tail instability, and it is thought that this is the mechanism in which they are produced in Type III bursts.

In the case of meteors, it is possible that an electron beam streaming through the ionosphere could interact with the meteor trail, or perhaps a beam could be generated within meteor trail itself. To explore the idea, let us consider an electron beam with a Maxwellian distribution traveling through a meteor trail, and lets assume the trail and beam have the same thermal velocity. Also for simplicity let us ignore the diffusion effects of the geomagnetic field. Under the weak growth rate approximation for a thermal plasma the growth rate \citep{Gurnett05} is given by:

\begin{equation}
\gamma = \frac{{\pi}}{2} \frac{\omega_{p}^{3}}{k^{2}} \frac{n_b}{n_e} \left. \frac{\partial F_0}{\partial v_z} \right|_{v_{z} = \omega / k},
\end{equation}


\noindent where $\omega_{p}  = 2 \pi f_{p}$, $k$ is the wave number, $n_{b}$ is the beam density, $n_{e}$ is the local electron density in the meteor trail, $v_z$ are the electron velocities in the direction of the beam, and $F_0$ is the normalized distribution function of the meteor plasma and the beam. Since this is only valid for large phase velocities, we only need to consider the contribution to $F_0$ by the beam, which we assume has a velocity at least four times the average thermal velocity. $F_b$ is represented by:

\begin{equation}
F_b = \frac{1}{\sqrt{\pi}v_{th}} e^{-\frac{(v_z - v_b)^{2}}{v_{th}^{2}}},
\end{equation}

\noindent where $v_b$ is the beam velocity and $v_{th}$ is the electron thermal velocity. Substituting $F_b$ for $F_0$ in equation 7, taking the derivative, and evaluating at $v_{z} = \omega / k$, we can find the growth rate at any frequency. The maximum growth rate is at $\omega \approx k(v_b - v_{th})$, evaluating at this frequency, and setting $e^{1} \approx 3$ we get:

\begin{equation}
\gamma \approx \frac{\sqrt{\pi}}{3} \frac{n_b}{n_e} \frac{\omega_{p}^{3}}{(kv_{th})^{2}} .
\end{equation}

Finally using the approximation $k \approx \omega_{p}/v_b$, we obtain,

\begin{equation}
\gamma \approx \frac{\sqrt{\pi}}{3} \frac{n_b}{n_e} \left(\frac{v_{b}}{v_{th}}\right)^{2}\omega_{p} .
\end{equation}

Insisting that the growth rate is greater than the electron-neutral collision frequency, i.e. $\gamma > \gamma_{en}$, we can solve for $n_{b} / n_{e}$ given a beam velocity. Assuming a 0.4 eV electron beam ($v_{b} \sim 4v_{th}$), to excite waves within a plasma with a plasma frequency of 37 MHz would require a beam density of $n_{b} = 2,000$ cm$^{-3}$ or about 0.01\% of the local electron density (1.7$\times10^7$ cm$^{-3}$). Although speculative, it is not unreasonable to hypothesize that such a weak beam could be generated directly within the trail itself. \citet{Dimant09} predicts large currents flowing outside and within meteors trail due to the large potential differences in this region of the ionosphere. The high electron/neutral collision frequency ($\sim 3 \times 10^{5}$ Hz) may indeed suppress the formation of any self generated electron beam from this current. However a beam might form if the current were to heat a narrow channel of plasma to high temperatures along the length of the trail. The decreased collision frequency of the heated channel may allow electrons to be accelerated to high velocities and escape into the surrounding plasma.


It should also be noted that the bump-on-tail instability does not require a directional, current-like beam; any source of high energy non-thermal electrons would be sufficient. Optical persistent trains, known to occur from high velocity fireballs, display an infrared to extreme ultra-violet (EUV) glow in the meteor trail and last for seconds up to hours after the disappearance of the meteor \citep{Borovicka06}. The luminosity of long lasting persistent trains is thought to be driven by exothermic chemical reactions, although all of the the relevant relations have not yet been identified. It is hypothetically possible that the energetic processes involved or the resultant emitted photons could pump energy into some electrons creating a bump in the distribution function. A related process has been suggested as an explanation for suprathermal electrons detected in the lower E-region, where thermal electrons may gain energy through interactions with vibrationally excited N$_2$ molecules and possibly form an non-Maxwellian distribution \citep{Oyama11}.



\subsection{Langmuir Wave Emission}

As emitters of Langmuir waves Type II and III Solar bursts are natural analogs to our proposed model for meteor emission. Although it should be noted that a complete theory describing the radiation mechanism of Langmuir waves from Solar bursts has yet to be determined. Several mechanisms have been proposed such as wave-wave interactions \citep{Cairns85,Cairns88}, linear mode conversion \citep{Hinkel92,Kim13}, and antenna radiation \citep{Papadopoulos78,Malaspina12}.
 
Three dimensional models of meteors trails have shown that atmospheric winds can create high levels of turbulence \citep{Oppenheim15}. This turbulence then creates clumps of plasma along the trail, which is in contrast to the idea that the trail remains a pristine cylinder with a Gaussian-like electron density distribution. It is not known if this could help drive Langmuir waves, however it would easily create large density gradients, which may aid the emission process. Especially in a scenario such as antenna radiation, which requires density clumps of size scales similar to the wavelength of the emitted waves (5 to 10 m).

A related phenomenon, known as stimulated electromagnetic emissions (SEE), has been observed in the ionospheric F-region where Langmuir waves are driven by high power transmitters \citep{Leyser01}. These waves are stimulated around the center frequency of the transmitter and are subsequently converted into electromagnetic waves, which have been observed to persist for milliseconds up to tens of seconds. While the Langmuir waves are generated artificially, the emission mechanism may be relatable to that in meteor trails.

\subsection{Spectral Sweeps}

The spectral sweeps of FB1 described in \S 4.2 appear distinct among the rest of the emission. The short burst-like nature and polarization do indeed set it apart from the bulk of the emission in both FB1 and FB2. While we do not have an explanation for the linear polarization we suggest that the frequency-time dispersion could be the result of high density clumps of plasma expanding into a lower density trail. 

To explore this idea, let us assume that small spheres of plasma with high electron density form within a less dense trail and diffuse isotropically in three dimensions. Let us also assume no significant recombination occurs on the time scale of the expansion. Using the conservation of particle number within an expanding sphere we obtain:

\begin{equation}
N_{i} = n_{ei} \frac{4}{3}\pi r_{i}^{3} = n_{e}(t) \frac{4}{3}\pi r^{3}(t),
\end{equation}

\noindent where $N$ is the initial number of free electrons, $n_{e}$ is the electron density, $r$ is the radius of the sphere, and the subscript $i$ denotes initial values. 

In this scenario we assume that the number of electrons contained within a growing sphere is constant, and that the entire sphere is kept at a uniform, yet decreasing electron density.  We can approximate the radius of the sphere with the mean square displacement 

\begin{equation}
r^{2}(t) = 6Dt + r_{i}^{2},
\end{equation}

\noindent where $D$ is the diffusion coefficient, which is typically given in cm$^{2}$s$^{-1}$, and depends exponentially on the height within the atmosphere \citep{Ceplecha98,McKinley61}. It should be noted that diffusion at high altitudes ($\gtrsim$ 100 km) becomes nonlinear and no longer follows the mean square displacement \citep{Dyrud01,Oppenheim03}. Moreover at altitudes above 90 km the trails display polarized diffusion along the geomagnetic field lines \citep{Jones91,Oppenheim15}. Both of these effects are subjects of current research, and further study of the radio emission may prove useful in the future of this field.

To avoid a more complicated diffusion scenario we assume that these clumps are lower than 95 km and that diffusion is unaffected by the geomagnetic field. Substituting the mean square displacement for the radius in equation 11 and substituting in the plasma frequency we get:

\begin{equation}
f_{p}^{2}(t) (6Dt + r_{i}^{2})^{3/2} = f_{pi}^{2} r_{i}^{3}.
\end{equation}

Solving for $t$ as a function of $f_{p}$ we get:

\begin{equation}
t(f_{p}) = \frac{r_{i}^{2}}{6D}\left[ \frac{f_{pi}^{4/3}}{f_{p}^{4/3}} - 1\right] + \tau,
\end{equation}

\noindent where $\tau$ is the time when the sweep began, $t(f_{pi})$. Given this model we refit the eight sweeps of B2 using Equation 3 with $b$ fixed to 4/3. The results are given in Table 2. We also fit the most defined sweep of B3, which is shown in Table 2 as \#9. Holding $b$ at 4/3 had only a small effect on the goodness of the fit with only minor decreases in the coefficient of determination (R$^{2}$). Despite this decrease, it is possible that the functional fit is the correct form, with the discrepancy being a result of a systematic error in the way the points were acquired. Nevertheless, if we assume that the fit is correct, the values for $a$ allow us to estimate $D$ using:

\begin{equation}
D = \frac{f_{pi}^{4/3} r_{i}^{2}}{6 a}.
\end{equation}

While we do not know the exact values for $r_{i}$ and $f_{pi}$, if the model is correct, ballpark estimates should get us to the right order of magnitude of $D$. If we assume $r_{i}$ = 400 cm and $f_{pi}\sim 100$ MHz, $D$ ranges from 9,000 to 11,000 cm$^{2}$ s$^{-1}$ for the eight modeled sweeps in B2 and 3,700 cm$^{2}$ s$^{-1}$ for the sweep seen in B3. The estimates of $D$ for all modeled sweeps are tabulated in Table 2.  

From the NASA All-Sky Fireball Network we know that the plasma trail for FB1 had an upper altitude of 100 km and lower altitude of 86 km. Based on the atmospheric density, we would expect meteors in this range to have dispersion coefficients in the range of 5,000 to 15,000 cm$^{2}$ s$^{-1}$ \citep{McKinley61,Ceplecha98,Wislez95}. Our estimates for the B2 sweep indicate they occurred in the middle ($\sim$ 92 km) of the trail, which is where the PASI images show B2 was pointed. This altitude also agrees with our assumption that the clumps were below 95 km, above which our diffusion assumptions become invalid. Our estimate for the B3 sweep indicates that it occurred just below the lower end ($\sim$ 85 km) of the trail, which is where the PASI images show B3 was pointed.  The fact that the estimate for the B3 sweep is slightly low may mean that we are under estimating the initial radius or initial plasma frequency for this sweep. Indeed the fact that these pulses are much broader in time and frequency than those seen in B2, could be a clue that they were larger in size to begin with.

At the same time we should emphasize that we really only guessed at the initial radius in the first place, and due to the square dependance, a small error in $r_{i}$ would have a large effect on $D$. Furthermore $r_{i}$ would most likely have a strong dependance on the height, further complicating the determination of $D$, which itself was more complicated than the simple linear model we assumed to begin with.

Next let us look at the modeled values for $c$. For all eight sweeps of B2 the values lay between 9.9 and 10.1 s. We can use Equations 3 and 8 to get:

\begin{equation}
\tau = a/f_{pi}^{4/3} +c
\end{equation}

Therefore assuming this model is correct, we know that all of these sweeps must have started sometime after $\sim$9.9 s. This implies that there must be some mechanism triggering the formation of these clumps long after the meteor disintegrated. Moreover it is interesting to note that the one sweep modeled in B3 had  $c$ = -6.7. Since the sweep could not start before the onset of the fireball,  we set $\tau \geq 0$ s, which yields $f_{pi} \leq$ 106 MHz. This limit  is certainly well within the physical bounds, and this would mean that $r_{i} \gtrsim$ 500 cm for the sweeps in B3, assuming that $D \gtrsim$ 3,700 cm$^{2}$ s$^{-1}$.

The regular periodicity of the sweeps could be a sign that the plasma is interacting with a large amplitude acoustic wave with a period of 0.1 to 0.2 s (5 to 10 Hz). Such a wave may be resonating within the trail and forming the dense plasma clumps through large neutral density fluctuations and plasma instability. These clumps would then begin to diffuse outward, sweeping through plasma frequency with a temporal separation of 0.1 to 0.2 s. A formal derivation of this mechanism is beyond the scope of this paper, but future exploration of this idea could prove useful. 

As noted earlier we did not take into account the effects from the geomagnetic field, which decreases the diffusion rate in the direction perpendicular to the geomagnetic field. At sufficiently high altitudes ($\gtrsim$ 100 km) the diffusion is much greater in the direction of the geomagnetic field, creating a non-spherically symmetric expansion \citep{Ceplecha98,Oppenheim15}. The small clumps of plasma would then expand as an ellipsoid rather than a sphere, meaning the functional dependence of $t$ on $f_p$ would be different. Similar to the derivation of Equation 14, it can be shown that in the case of a cylinder expanding only in length, $t(f_p) \propto f_{p}^{-2}$. Therefore an expanding ellipsoid would have a value of $b$ between 4/3 and 2. 

In addition it should be noted that most large fireball trails have a dusty component \citep{Kelley98}, and it is thought that a build up of charge onto the dust may aid in the process of long duration meteor echoes \citep{Kelley03}. We did not consider this dusty component but it may indeed need to be considered in further studies of the radio emission.

\section{Conclusions}

The dynamic spectra presented here confirm the earlier discovery of long duration, non-thermal radio emission from fireballs, occurring near the plasma frequency. The spectra also confirm that the bulk of the emission is broadband, unpolarized and has a fairly smooth and steep spectrum. Furthermore the high temporal and frequency resolution of the LWA1 has allowed for the detection of finely structured, linearly polarized sweeps on top of the smooth unpolarized emission in one of the two fireballs observed.

Using the all sky imaging capabilities of the LWA1 as well as the triangulation of the NASA All Sky Fireball Network, we correlated the different appearance of these sweeps with different heights of the plasma trail. The broad sweeps of $\sim$ 1 s width and higher dispersion are connected with the lower end of the trail at a height of $\sim$ 85 km, and the narrow sweeps of $\sim$ 0.2 s width with lower dispersion have been connected with the middle portion of the trail at a height of $\sim$ 92 km. It was shown that the bulk of the dispersion cannot be caused by typical electromagnetic dispersion within a plasma and that some other explanation is necessary. The sweeps also displayed complicated frequency dependence in Stokes Q with 80\% -Q at 43 MHz and 50\% +Q at 51 MHz, and the amount of polarization dropping off significantly within 1 MHz on either side of these frequencies. It was shown that this change in polarization angle is almost certainly a feature of the emission and not due to Faraday rotation. 

We postulated that the emission is most likely due to radiation of simple plasma oscillations (Langmuir waves), and modeled the sweeps as clumps of high density plasma diffusing into a lower density surrounding. The expanding clump model does provide a reasonable explanation for the frequency-time dispersion as the plasma frequency of the clump changes. However this model does not explain the observed linear polarization features.

A second LWA station is under construction at the Sevileta National Wildlife Refuge, about 60 km distant from LWA1. This should allow for triangulation of fireballs  leading to precise measurements of the heights and trajectories of the fireballs.  Eventually it will be possible to employ radio interferometry techniques to image the fireballs in detail and measure their expansion.

\begin{acknowledgments}
We thank the anonymous referees for thoughtful comments.

Construction of the LWA1 has been supported by the Office of Naval Research under Contract N00014-07-C-0147. Support for operations and continuing development of the LWA1 is provided by the National Science Foundation under grants AST-1139963 and AST-1139974 of the University Radio Observatory program. 

All of the beam formed data in this article is publicly available at the LWA1 Data Archive (lda10g.alliance.unm.edu). All of the PASI image data is available by request from Kenneth Obenberger (kso1987@unm.edu).

\end{acknowledgments}

\end{article}

\clearpage

\begin{table}

\caption{Pulse Sweep Parameters}
\centering

\begin{tabular}{|c|c|c|c|c|}
\hline
Sweep \#		&	$a$		&	$b$		&	$c$	 &		R$^{2}$\\
 	 		&	s MHz$^{b}$		&	 			&	s			&	  \\
\hline
1			&	1042			&	1.31			&	10.0				&	0.9993\\
\hline
2			&	1778			&	1.48 			&	10.9				&	0.9994\\
\hline
3			&	3413			&	1.67			&	11.6				&	0.9996\\
\hline
4			&	5630			&	1.81			&	12.0				&	0.9998\\
\hline
5			&	836			&	1.19			&	9.0				&	0.9997\\	
\hline
6			&	608			&	1.07	 		&	8.0				&	0.9996\\
\hline
7			&	493			&	0.989		&	7.1				&	0.9995\\
\hline
8			&	280			&	0.7746		&	4.0				&	0.9998\\
\hline
\end{tabular}										
\tablenotetext{a}{Parameters $a$, $b$ and $c$ are free. }

\end{table}

\begin{table}

\caption{Pulse Sweep Parameters}
\centering

\begin{tabular}{|c|c|c|c|c|c|}
\hline
Sweep \#		&	$a$		&	$b$		&	$c$	 &		R$^{2}$	&	$D$\\
 	 		&	s MHz$^{4/3}$		&	 			&	s			&	 	  &	 cm$^{2}$ s$^{-1}$ \\
\hline
1			&	1,101			&	4/3  			&	10.1				&	0.9993		&	11,242\\
\hline
2			&	1,120			&	4/3			&	10.2				&	0.9993		&	11,051\\
\hline
3			&	1,158			&	4/3			&	10.2				&	0.9995		&	10,689\\
\hline
4			&	1,222			&	4/3			&	10.0				&	0.9995		&	10,129\\
\hline
5			&	1,274			&	4/3			&	9.9				&	0.9997		&	9,716\\	
\hline
6			&	1,315			&	4/3 			&	9.9				&	0.9995		&	9,413\\
\hline
7			&	1,356			&	4/3			&	9.9				&	0.9994		&	9,128\\
\hline	
8			&	1,382			&	4/3			&	9.9				&	0.9997		&	8,956\\		
\hline	
9			&	3,344			&	4/3			&	-6.7				&	0.9990		&	3,701\\	
\hline
\end{tabular}										
\tablenotetext{a}{Parameters $a$, $b$ and $c$ are free. }

\end{table}

\clearpage

\begin{figure}
\noindent\includegraphics[width = 7in]{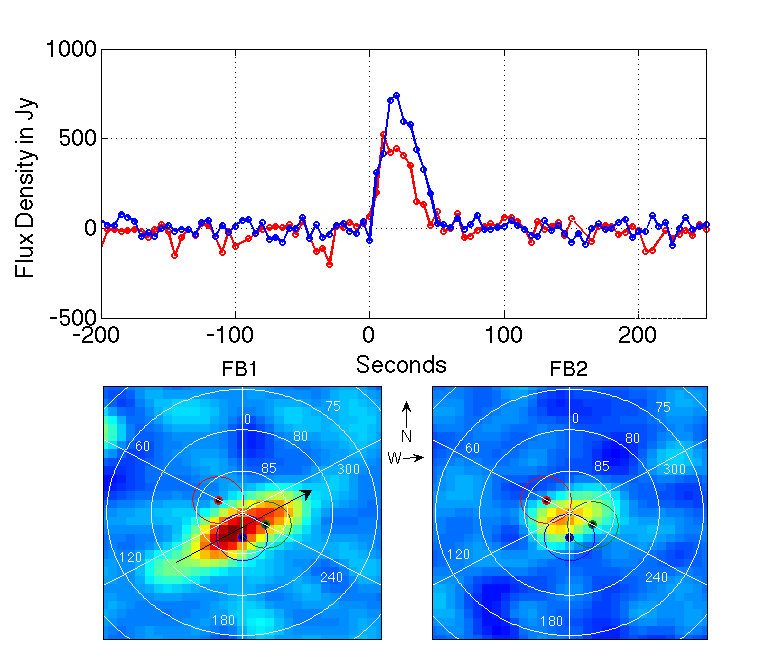}
\caption{Top: Light curves showing the flux density from the brightest region from both fireballs. Units are in Janskys. (1 Jy = 10$^{-23}$ erg s$^{-1}$ cm$^{-2}$ Hz$^{-1}$) Bottom: Closeups of the averaged images of the FB1 (left) FB2 (right). The image for FB1 also shows an arrow indicating the trajectory of the fireball heading NW. Also shown are the locations of B1 (Red), B2 (Blue), and B3 (green), each encircled by the estimated full width at half maximum (FWHM) beam size at 38 MHz. A horizontal coordinate map is overlaid on the image, and each pixel is approximately $1^{\circ} \times1^{\circ}$. }
\end{figure}

\begin{figure}
\noindent\includegraphics[width = 7in]{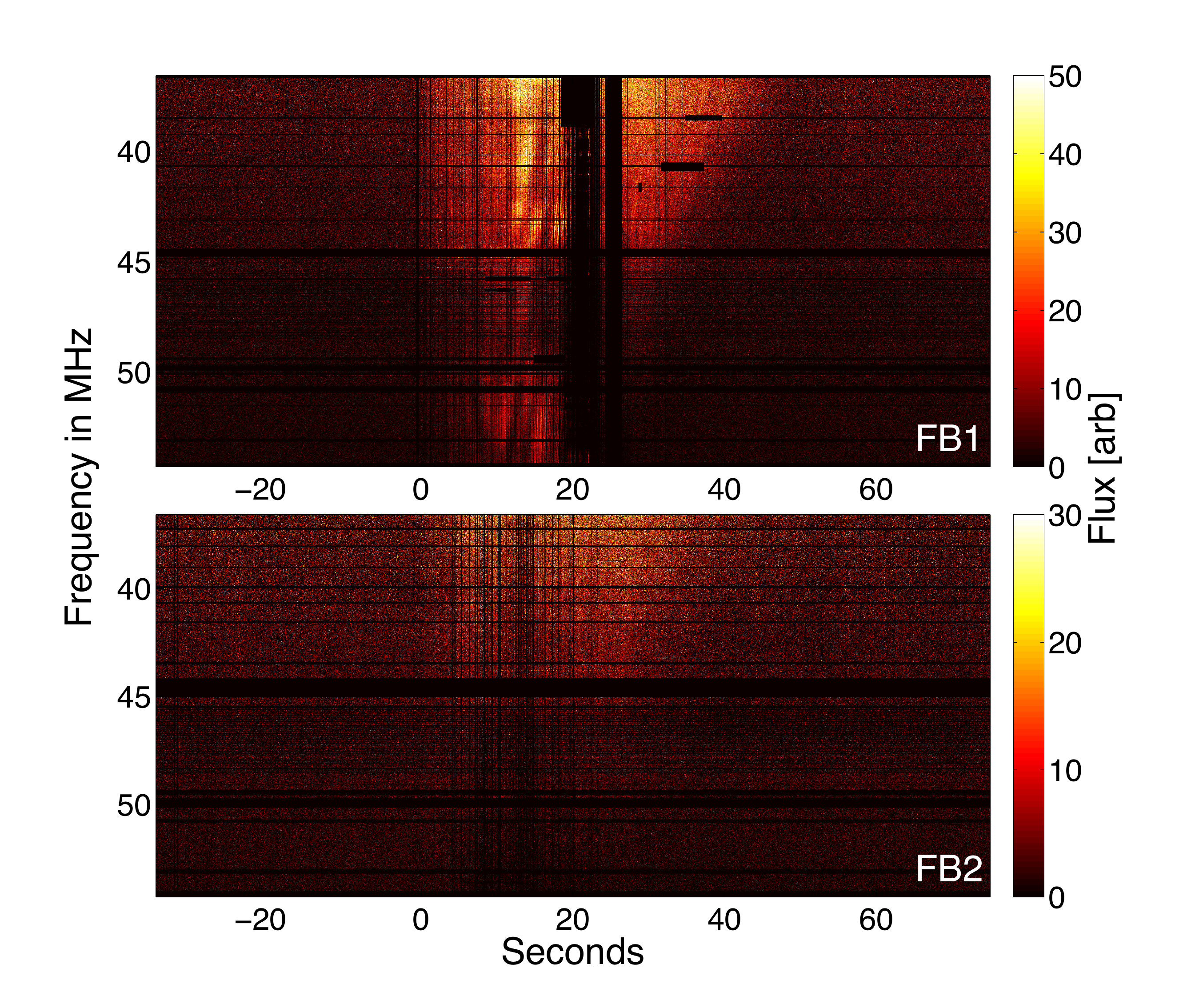}
\caption{Dynamic spectra from B2 for FB1 (top) and FB2 (Bottom). Flux scales are in arbitrary units, but are the same for both FB1 and FB2. However notice that the color axes of FB1 and FB2 are different.}
\end{figure}

\begin{figure}
\noindent\includegraphics[width = 7in]{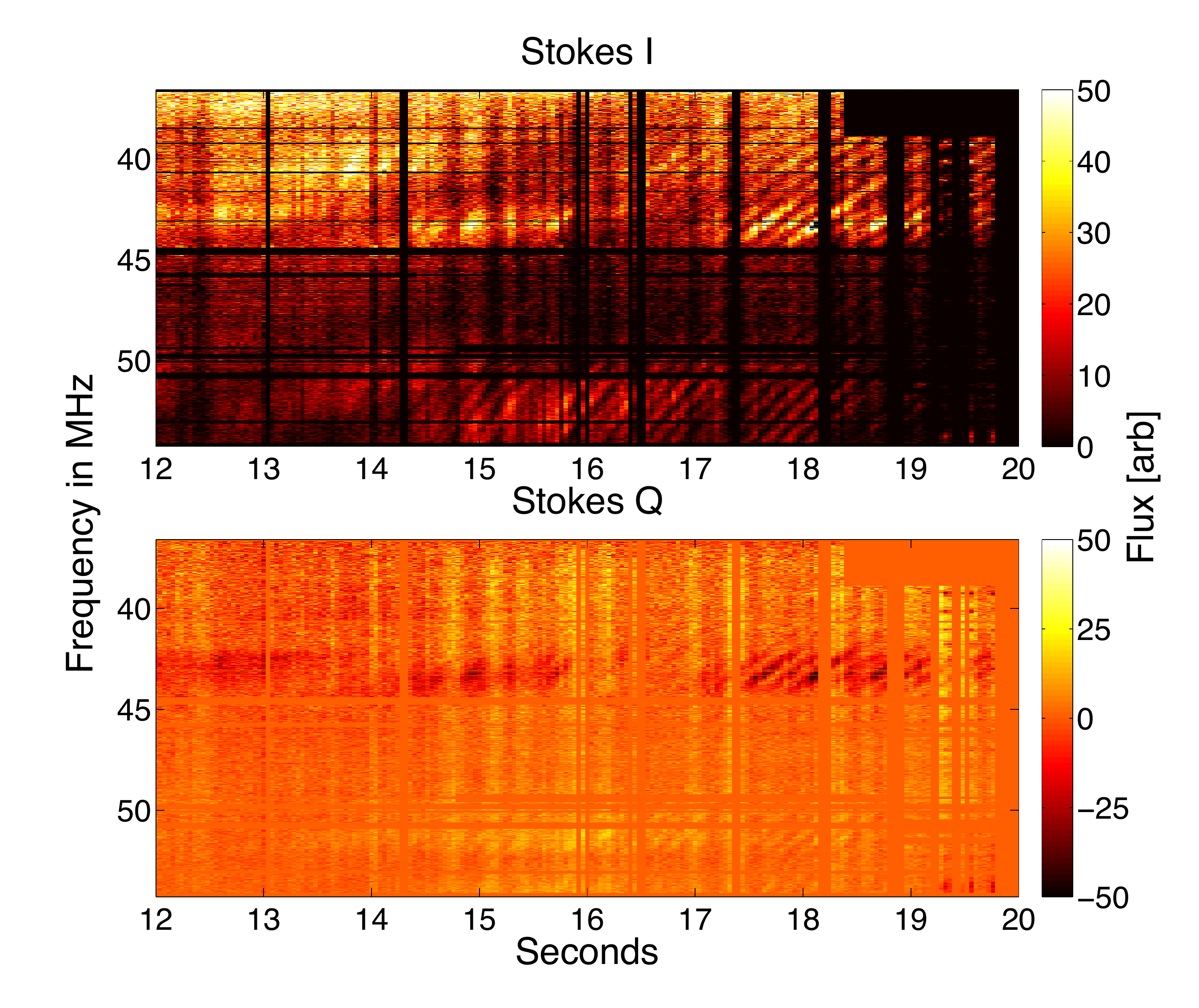}
\caption{A closeup of the dynamic spectra from B2 of FB1 from 12 to 20 seconds, shown in both Stokes I (top) and Stokes Q (bottom). For Stokes Q, positive values correspond to East-West polarization, and negative values correspond to North-South Polarizations. }
\end{figure}

\begin{figure}
\noindent\includegraphics[width = 7in]{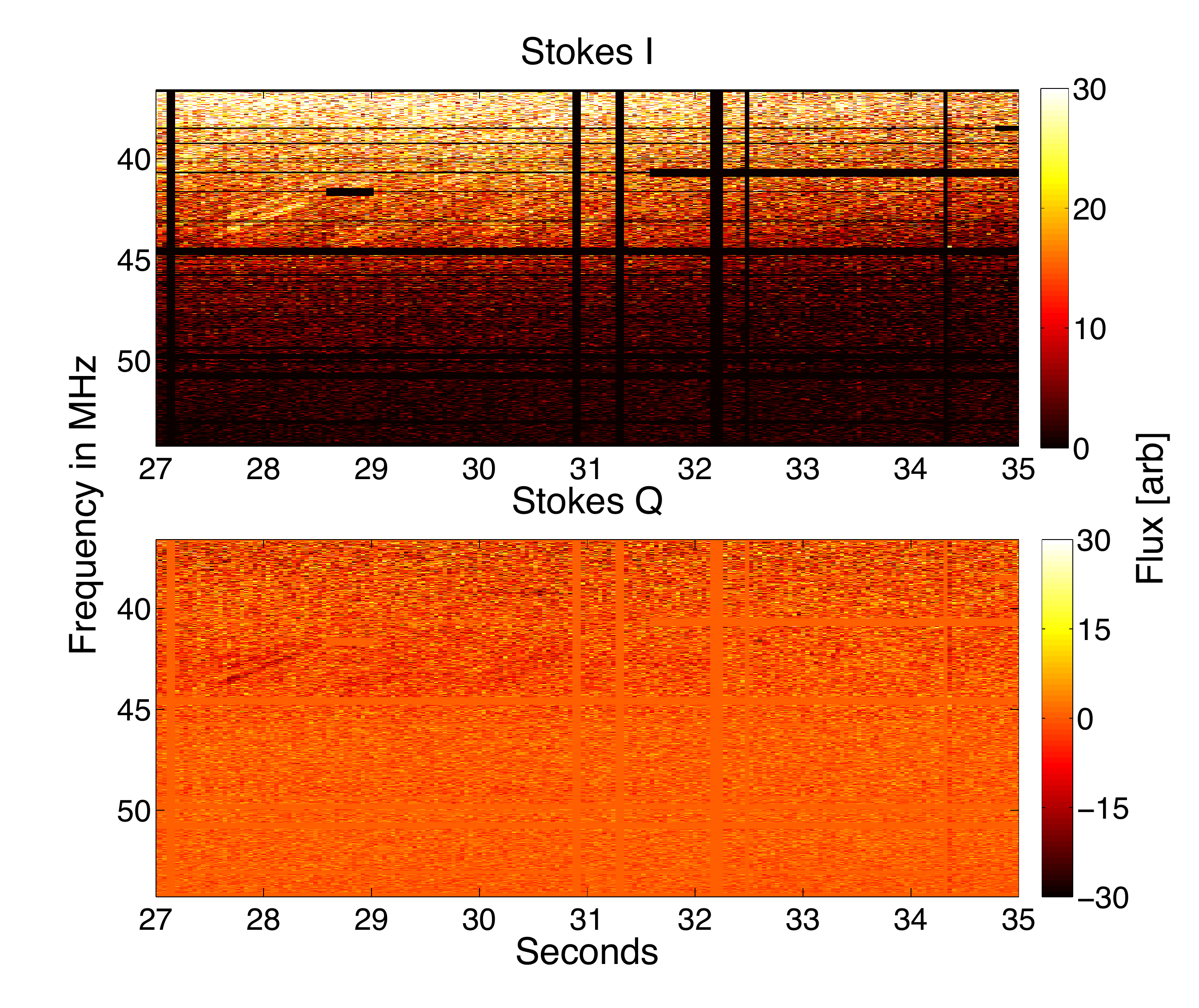}
\caption{A closeup of the dynamic spectra from B2 of FB1 from 27 to 35 seconds, shown in both Stokes I (top) and Stokes Q (bottom). For Stokes Q, positive values correspond to East-West polarization, and negative values correspond to North-South Polarizations.}
\end{figure}

\begin{figure}
\noindent\includegraphics[width = 7in]{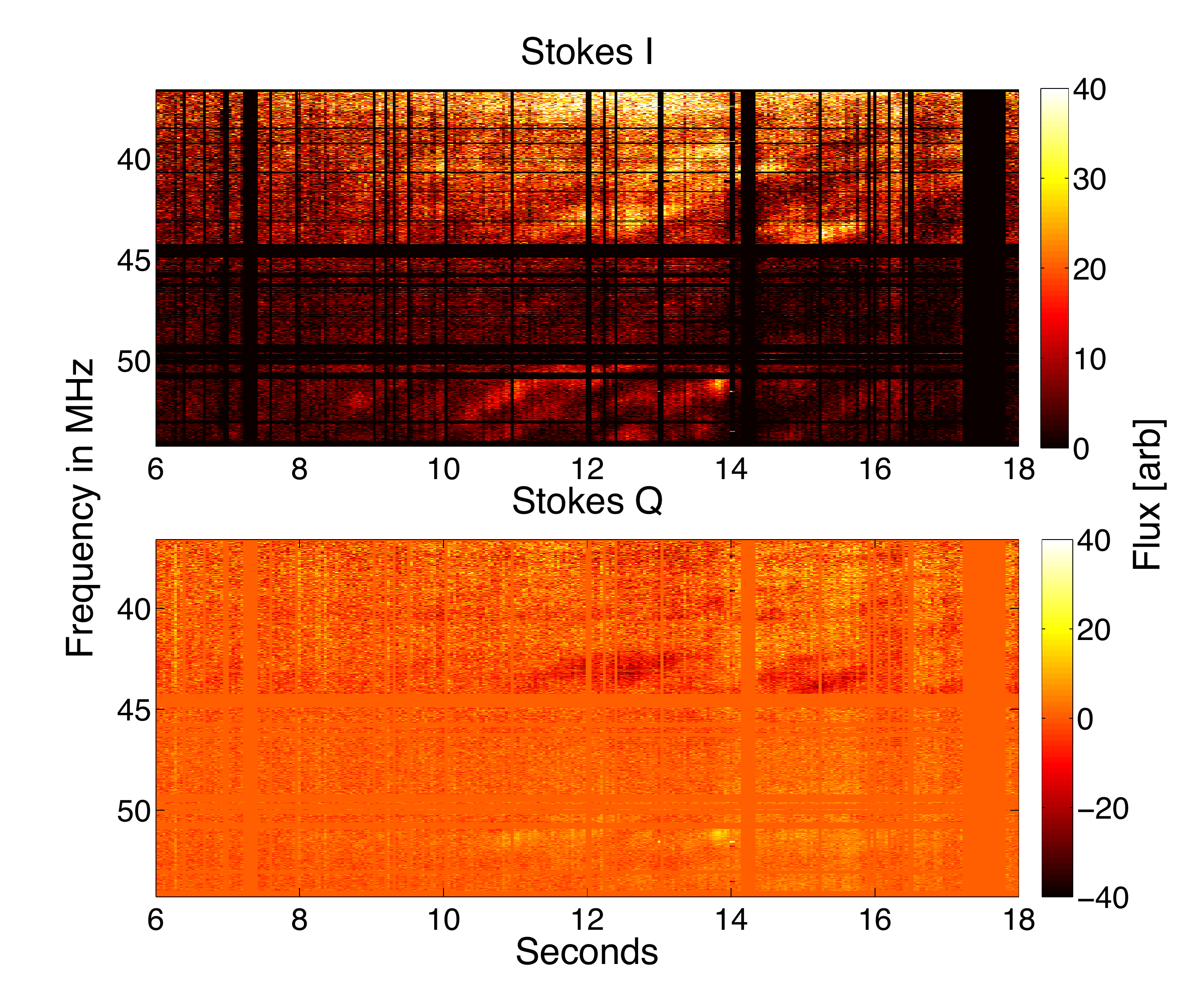}
\caption{A closeup of the dynamic spectra from B3 of FB1 from 6 to 18 seconds, shown in both Stokes I (top) and Stokes Q (bottom). For Stokes Q, positive values correspond to East-West polarization, and negative values correspond to North-South Polarizations. }
\end{figure}

\begin{figure}
\noindent\includegraphics[width = 7in]{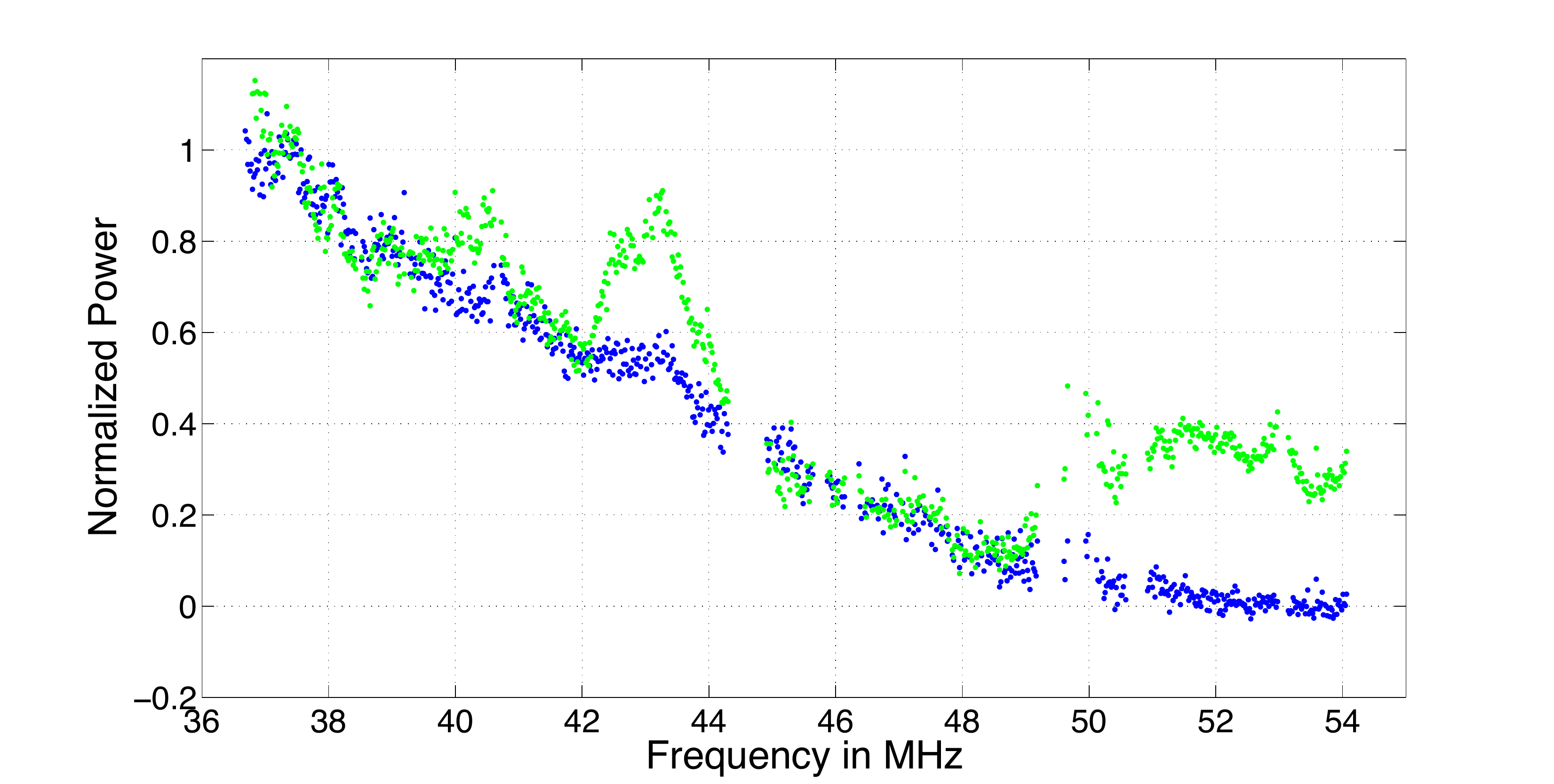}
\caption{Comparing the averaged spectra from B2 of the 14th to 20th second (green) and the 26th to 34th second (blue) for FB1. Both averaged spectra have been corrected for bandpass, sky, and beam contributions, and are normalized to the brightest averaged frequency. As can be seen there is enhanced emission from 40 to 41, 42 to 44, and 49 to 54 MHz. These bumps correspond to the polarized frequency sweeps. }
\end{figure}

\begin{figure}
\noindent\includegraphics[width = 7in]{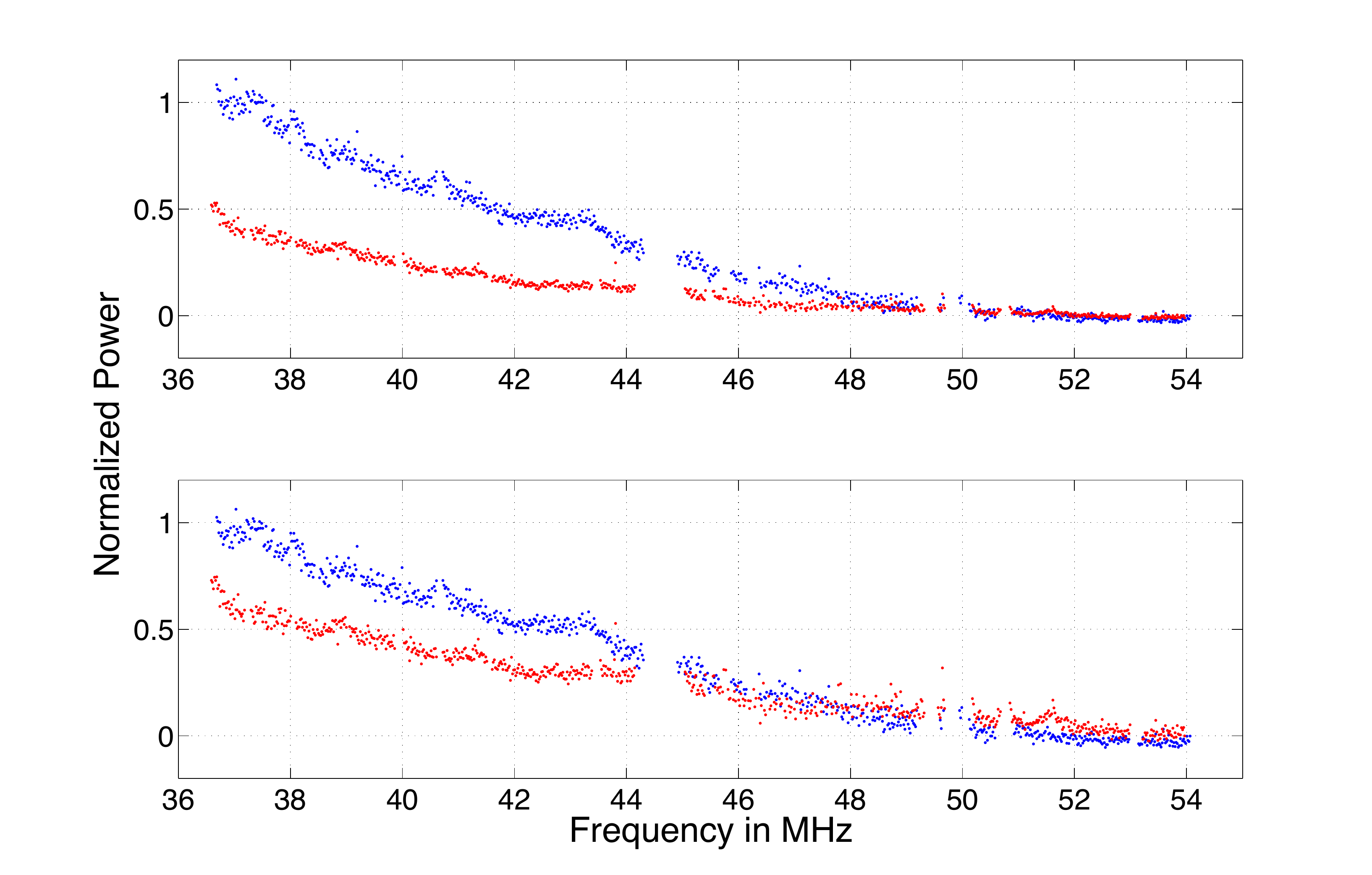}
\caption{(Top) Mean spectra from FB1 (blue) and FB2 (red) after bandpass and sky contribution are removed, but before the beam contribution is removed. (Bottom) Mean spectra after beam contribution has been divided out. Both plots are normalized to the brightest region of FB1. Note: For FB1 only the averaged spectra between the 26th and 34th seconds are used, whereas for FB2, the entire duration of the emission is averaged. }
\end{figure}


\end{document}